\documentstyle[12pt]{article}
\def\be{\begin{eqnarray}}
\def\ee{\end{eqnarray}}
\def\ba{\begin{array}}
\def\ea{\end{array}}
\begin{document}
\begin{center}
{\LARGE   { Generation of heterotic string theory solutions\\
\vskip 3mm
from the stationary Einstein-Maxwell fields}}
\end{center}
\vskip 25mm
\begin{center}
{\bf \large {Oleg V. Kechkin}}
\end{center}
\begin{center}
DEPNI, Institute of Nuclear Physics\\
M.V. Lomonosov Moscow State University\\
119899 Moscow, Vorob'yovy Gory, Russia\\
e-mail:\, kechkin@depni.npi.msu.su
\end{center}
\vskip 25mm
\begin{abstract}
A new formalism for construction of the stationary solutions
is developed for the four-dimensional gravity coupled to the
dilaton, Kalb-Ramond and two Maxwell fields in a low-energy
heterotic string theory form. The result of generation is
automatically invariant in respect to subgroup of the
stationary charging symmetry transformations; the generation
can be started from the stationary Einstein-Maxwell fields.
The formalism is given both in real and new compact complex
form, the result of maximal symmetry extension of the
stationary Einstein-Maxwell theory to discussing string
gravity model is explicitly written down.
\end{abstract}
\vskip 10mm
\begin{center}
\end{center}
\renewcommand{\theequation}{\thesection.\arabic{equation}}
\newpage
\section{Introduction}

String theory leads to field theories of gravity coupled to
different matter fields in its low energy limit
\cite{Kir}.
For the heterotic string theory case these matter fields
include a set of Abelian vector ones, so the corresponding
gravity model
can be considered as some string theory generalization of the
classical Einstein-Maxwell theory \cite{EMT}.

The heterotic string theory motivated gravity model possesses a
rich system of on-shell symmetries \cite{MahSch}, \cite{Sen4},
\cite{Sen3}, \cite{BKR}, \cite{HK1}, \cite{HK2}. These
symmetries can be used for generation of new solutions from the
known ones accordingly the classical Sofus Lee approach. To apply
the symmetry transformation technique one must find some
special truncation of the studying theory which in fact is the
system with well known solution space. In this case it begins
possible to
construct the symmetry extension of this subsystem to the whole
theory. Such a program had been extensively realized in the
heterotic string theory framework for the choice of stationary
General Relativity as the starting subsystem \cite{Youm},
\cite{OUR}.

In this article the stationary Einstein-Maxwell theory will put
down to work. This step is nontrivial; to clarify the question
let us consider the stringy dilaton-axion generalization of
Einstein-Maxwell theory. The corresponding action includes the
terms $e^{\alpha\Phi}F^2$ and $\beta\kappa F\tilde F$, where $F$
is the Maxwell stress tensor, $\Phi$ and $\kappa$ are the dilaton
and Pecci-Quinn axion fields, whereas $\alpha$ and $\beta$ are the
dilaton-Maxwell and axion-Maxwell constant couplings. For the
realistic string theory motivated gravity models
$\alpha,\,\beta\neq 0$, so if one puts $\Phi=\kappa=0$ on shell to
obtain the Einstein-Maxwell theory from the discussing model, one
additionally obtains the restrictions
\be\label{e0}
F^2=F\tilde F=0
\ee
from the original `$\Phi$' and `$\kappa$'-equations.
These restrictions leave only the plane-wave special case for the
possibility of string theory generalization of the Einstein-Maxwell
theory which is not sufficient for the practice. However, as it
will be shown below, it exists another form of enclosure of
Einstein-Maxwell theory into the heterotic string gravity models
which is free of any restrictions. This new approach is directly
related to some new formalism based on the use of matrix Ernst-type
potentials which had been developed for the general heterotic
string gravity models in \cite{HK2}, \cite{OK}. Below it will be
explored for the total extension of the stationary Einstein-Maxwell
theory to the heterotic string theory case in respect to its subgroup
of the stationary charging symmetry transformations. The developed
formalism provides a compact and convenient tool for the generation
of charged and asymptotically flat solutions in the low-energy
heterotic string theory.

\section{New matrix formalism and charging symmetries}
\setcounter{equation}{0}
Let us now formulate the heterotic string gravity model under
consideration. The starting effective field theory lives in the
$(d+3)$-dimensional space-time with the signature
$+\cdots+;-,+,+$ and describes the bosonic zero-mass modes of
excitation of the heterotic string theory: the dilaton $\Phi$,
Kalb-Ramond field $B_{MN}$ ($M,N=1,...,d+3$), $n$ Abelian fields
$A^I_M$ \, ($I=1,...,n$) and the metric $G_{MN}$. The corresponding
action is \cite{Kir}:
\be\label{e1}
S_{d+3}=\int d X \sqrt{-{\rm det}G_{MN}}\,e^{-\Phi}
\left ( R_{d+3}+\Phi_{,M}\Phi^{,M}-\frac{1}{12}H_{MNK}H^{MNK}-
\frac{1}{4}F^I_{MN}F^{I\,MN}\right ),
\ee
where $H_{MNK}=B_{NK,M}-1/2\,F_{NK}^IA_M^I+{\rm cyclic}\,\,
(M,N,K)$ and $F_{MN}^I=A_{N,M}^I-A_{M,N}^I$.
Following \cite{MahSch},
\cite{Sen4},\cite{Sen3} we consider a toroidal compactification of the
first $d$ dimensions. In this case all the fields begin independent on
the coordinates $X^m$ with $m=1,...,d$ and depend on the ones
$x^{\mu}=X^{d+\mu}$, \,\, $\mu=1,2,3$. The resulting $3$--dimensional
system is equivalent on-shell to some concrete nonlinear $\sigma$-model
coupled to gravity \cite{Sen3}. Such $\sigma$-models which parametrize
a coset space had been classified in \cite{BGM}; for the theory under
consideration the coset is $O(d+1,d+1+n)/O(d+1)\times O(d+1+n)$
\cite{Sen3}. The resulting $3$-dimensional gravity model describes the
set of $(d+1)\times(d+1+n)$ functionally independent scalar fields
coupled to the effective $3$-metric $h_{\mu\nu}$. It can be alternatively
described in terms of the null-curvature $[2(d+1)+n]\times [2(d+1)+n]$
matrix ${\cal M}$ \cite{Sen3},\cite{HK1}, the pair of matrix Ernst
potentials ${\cal X}$ and ${\cal A}$ of the dimension $(d+1)\times(d+1)$
and $(d+1)\times n$ \cite{HK1}, and also using the
$(d+1)\times(d+1+n)$-dimensional matrix potential ${\cal Z}$ \cite{HK2},
\cite{OK}. In Appendix A one can find the definition of $\sigma$-model
scalar fields and the relations between these three alternative matrix
formulations.

For our purposes we need in the ${\cal Z}$-formulation; the effective
$3$-dimensional action reads \cite{OK}:
\be\label{e2}
S_3=\int dx \left \{-R_3+{\rm Tr}\left [ \nabla {\cal Z}
\left ( \Xi-{\cal Z}^T\Sigma{\cal Z}\right )^{-1}\nabla Z^T
\left ( \Sigma-{\cal Z}\Xi{\cal Z}^T\right )^{-1}
\right ]\right \},
\ee
where $\Sigma$ and $\Xi$ are $(d+1)\times(d+1)$ and $(d+1+n)\times(d+1+n)$
matrices of the form ${\rm diag}(-1,-1;1,...,1)$. In \cite{OK} one can
find full information about this representation; here we need only in
trivial fact that the transformation
\be\label{e3}
{\cal Z}\rightarrow{\cal C}_L{\cal Z}{\cal C}_R
\ee
is a symmetry of the action (\ref{e2}) if
\be\label{e4}
{\cal C}_L^T\Sigma{\cal C}_L=\Sigma, \quad {\cal C}_R^T\Xi{\cal C}_R=\Xi,
\ee
i.e. if ${\cal C}_L\in O(2,d-1)$ and ${\cal C}_R\in O(2,d-1+n)$. In
\cite{HK2} it is shown that this symmetry is equivalent to the charging
symmetry subgroup of the complete group of 3-dimensional symmetries of the
theory. The transformations from the charging symmetry subgroup leave
unchanged the trivial solution ${\cal Z}=0, \, h_{\mu\nu}=\delta_{\mu\nu}$
of motion equations corresponding the action (\ref{e2}). This trivial
$3$-dimensional solution, being expressed in terms of the original
multidimensional string theory fields, arises in framework of the
asymptotically flat solutions or solutions possessing the nontrivial NUT,
magnetic and other similar charges (see \cite{EMT} for the Einstein-Maxwell
analogies). The asymptotically flat solutions play extremely important role
in physical interpretations of the theory and often (for example, in the
black hole physics \cite{Youm}, \cite{CH}) the charging symmetry subgroup
is the only resultative
part of the complete group of symmetries. The remaining part of
symmetries provide the nonimportant shift of values of the physical fields
at the spatial infinity.

\section{Generation using real potentials}
\setcounter{equation}{0}
Now let us consider the special case of $d=1$, $n=2$, which is also interesting
in framework of the $D=N=4$ supergravity \cite{Kal}. Also let us consider the
charging symmetry transformations which can be continuously related to the
identical one, i.e. let ${\cal C}_L\in SO(2)$, ${\cal C}_R\in SO(2,2)$ in the
following consideration. The nearest goal is to write down these transformation
matrices in the explicit form, which can be obtained through exponentiation of the
corresponding infinithesimal transformations.

So, let $\hat\lambda_L$ and $\hat\lambda_R$ are the general $so(2)$ and $so(2,2)$
algebra matrices, i.e. for the group matrices one has
\be\label{e5}
{\cal C}_L=\exp{\hat\lambda_L},\qquad {\cal C}_R=\exp{\hat\lambda_R}.
\ee
The algebraic matrices can be calculated as the general linear combinations of the
corresponding generators,
\be\label{e6}
\hat\lambda_L=\lambda_0\Gamma_0,\qquad \hat\lambda_R=\lambda_K\Gamma_K,
\ee
where the generators satisfy the algebraic relations
\be\label{e7}
\Gamma_0^T=-\Gamma_0,\qquad \Gamma_K^T=-\Sigma_3\Gamma_K\Sigma_3,
\ee
with
\be\label{e8}
\Sigma_3=\left (\ba{cc}
1&0\cr
0&-1
\ea\right ).
\ee
The generator $\Gamma_0$ is the $2\times 2$ matrix, whereas $\Gamma_K$ are the
$4\times 4$ ones; they can be taken in the following form:
\be\label{e9}
\Gamma_0&=&\epsilon=-i\sigma_2;\nonumber\\
\Gamma_1^1&=&
\left (\ba{cc}
0&1\cr
1&0
\ea\right ), \quad
\Gamma_2^1=
\left (\ba{cc}
0&-\epsilon\cr
\epsilon&0
\ea\right ), \quad
\Gamma_3^1=
\left (\ba{cc}
\epsilon&0\cr
0&-\epsilon
\ea\right );\nonumber\\
\Gamma_1^2&=&
\left (\ba{cc}
0&\sigma_1\cr
\sigma_1&0
\ea\right ), \quad
\Gamma_2^2=
\left (\ba{cc}
0&\sigma_3\cr
\sigma_3&0
\ea\right ), \quad
\Gamma_3^2=
\left (\ba{cc}
\epsilon&0\cr
0&\epsilon
\ea\right ),
\ee
where the usual Pauli matrices have been used (in our notations
$\{\Gamma_K\}=\{\Gamma^1_{\mu},\Gamma^2_{\nu}\}$). Then, it is easy to
verify that
\be\label{e10}
[\Gamma^1_{\mu},\Gamma^2_{\nu}]=0
\ee
and that the following multiplication relations take place:
\be\label{e11}
\left (\Gamma^a_1\right )^2&=&\left (\Gamma^a_2\right )^2=
=-\left (\Gamma^a_3\right )^2=1,\nonumber\\
\Gamma^a_1\Gamma^a_2&=&-\Gamma^a_2\Gamma^a_1=\Gamma^a_3,\nonumber\\
\Gamma^a_2\Gamma^a_3&=&-\Gamma^a_3\Gamma^a_2=-\Gamma^a_1,\nonumber\\
\Gamma^a_3\Gamma^a_1&=&-\Gamma^a_1\Gamma^a_3=-\Gamma^a_2,
\ee
where $a=1,2$. Thus, $\hat\lambda_R=\hat\lambda^1_R+\hat\lambda^2_R$,
where $\hat\lambda^a_R=\lambda^a_{\mu}\Gamma^a_{\mu}$, and from Eqs.
(\ref{e10}), (\ref{e11}) it follows that the matrices $\hat\lambda^a$
parametrize two commuting $4\times 4$ realizations of the $so(2,1)$
algebra. This fact is related to the isomorphism $so(2,2)\sim so(2,1)
\bigoplus so(2,1)$ and will crucial for our analysis. Using Eqs.
(\ref{e10}) and (\ref{e11}) it is possible to calculate
${\cal C}_R$; the result reads:
\be\label{e12}
{\cal C}_R={\cal C}^1_R{\cal C}^2_R={\cal C}^2_R{\cal C}^1_R,
\ee
where
\be\label{e13}
{\cal C}^a_R=\cosh\lambda^a+\frac{\sinh{\lambda^a}}{\lambda^a}
\tilde\lambda^a
\ee
and $\lambda^a=\{(\lambda^a_1)^2+(\lambda^a_2)^2-(\lambda^a_3)^2\}^{1/2}$.
For the matrix ${\cal C}_L$ one has:
\be\label{e14}
{\cal C}_L=\cos{\lambda_0}+\sin{\lambda_0}\Gamma_0.
\ee
Finally, Eqs. (\ref{e3}), (\ref{e12}) (\ref{e13}) and (\ref{e14})
together with the definition of the potential ${\cal Z}$, given in
Appendix A, completely define the generation using continuous charging
symmetries in this theory in the explicit form.

Now let us define the starting subsystem for the generation procedure.
To do it, let us put ${\cal Z}=({\cal Z}_1,\,{\cal Z}_2)$, where
${\cal Z}_a$ are the $2\times 2$ block components. Let us also put
\be\label{e15}
{\cal Z}_a=z_a^{'}-z_a^{''}\epsilon,
\ee
where
\be\label{e16}
z_a=z_a^{'}+iz_a^{''}
\ee
is the complex function. The first statement crucial
for the following consideration is that the pair $({\cal Z}, h_{\mu\nu})$
gives the solution of motion equations for the theory (\ref{e2}) if the pair
$({\bf z},h_{\mu\nu})$ is the solution for the theory
\be\label{e17}
S_{EM}=\int dx \left \{-R_3+
2\frac{\nabla {\bf z}\left ( \sigma_3-{\bf z}^+{\bf z}\right )^{-1}
\nabla {\bf z}^+}{1-{\bf z}\sigma_3{\bf z}^+}
\right \},
\ee
where
\be\label{e17'}
{\bf z}=(z_1,z_2)
\ee
and the relation (\ref{e15}) takes place. The proof is
related to the fact that the matrix (\ref{e15}) is the so called
`exact' realization of the complex function (\ref{e16}) and the multiplier `2'
is defined by the matrix dimensionality of this realization. Then, the
absence of any additional restrictions follows from the closing character
of complex number basis $(1,i)$. The second statement is that the theory
(\ref{e17}) coincides with the stationary Einstein-Maxwell theory. To prove
it, let us introduce the alternative complex variables
\be\label{e18}
{\cal E}=\frac{1-z_1}{1+z_1},\qquad {\cal F}=\frac{\sqrt 2z_2}{1+z_1}
\ee
(it is interesting to note that the inverse map $({\cal E}, {\cal F})
\rightarrow (z_1,z_2)$ has the same form as Eq. (\ref{e18})). Than, in terms
of these new potentials,
\be\label{e19}
S_{EM}=\int dx \left \{-R_3+
\frac{1}{2F^2}\left | \nabla {\cal E}+\bar{\cal F}\nabla{\cal F}\right |^2
-\frac{1}{f}\left | {\cal F}\right |^2
\right \},
\ee
where $f=1/2({\cal E}+\bar{\cal E}+|{\cal F}|^2)$, i.e. exactly the stationary
Einstein-Maxwell theory action \cite{EMT}, \cite{Maz} with ${\cal E}$ and
${\cal F}$ as Ernst potentials \cite{Ernst}.

The third important statement can be formulated as the nonimportance of the
charging symmetry subgroups given by the matrices ${\cal C}_L$ and ${\cal C}_R^1$
for generation starting from the stationary Einstein-Maxwell theory. Actually,
let us suppose that the solution $({\bf z},h_{\mu\nu})$ of the theory (\ref{e17})
is charging symmetry complete, i.e. the symmetry transformations preserving the
trivial ${\bf z}$-value had been applied in framework of the Einstein-Maxwell theory
itself. These transformations are obviously given by the single map
\be\label{e20}
{\bf z}\rightarrow {\bf C}_L{\bf z}{\bf C}_R^1
\ee
with ${\bf C}_L\in U(1)$ and ${\bf C}_R^1\in SU(1,1)$. It is easy to see that the
substitution (matrices $\rightarrow$ numbers)
\be\label{e21}
1\rightarrow 1,\qquad -\epsilon\rightarrow i
\ee
generates the map $\Gamma_0\rightarrow {\bf\Gamma}_0$, $\Gamma_{\mu}^1\rightarrow
{\bf\Gamma}_{\mu}$, where
\be\label{e22}
{\bf\Gamma}_0&=&i,\nonumber\\
{\bf\Gamma}_1&=&\sigma_1,\quad{\bf\Gamma}_2=\sigma_2\quad{\bf\Gamma}_3=i\sigma_3.
\ee
These generators belong to the $u(1)$ and $su(1,1)$ algebras
(${\bf\Gamma}_{\mu}^+=-\sigma_3{\bf\Gamma}_{\mu}\sigma_3$) and define
the corresponding group elements through the exponentials. These group elements can
be obtained from
the matrices $C_L$ and $C_R^1$ (see Eqs. (\ref{e13}), (\ref{e14})) using
the substitution (\ref{e21}) and the fact that ${\bf\Gamma}_{\mu}$ satisfy the same
multiplication relations as the ones given in Eq. (\ref{e11}). Thus, one can omit the
charging symmetry subgroup $SO(2)\times SO(2,1)\sim U(1)\times SU(1,1)$ realized
on the matrices $C_L$ and $C_R^1$ and forming a total group of charging symmetries
of Einstein-Maxwell theory if one starts in generation from the charging symmetry
complete solutions of the stationary Einstein-Maxwell theory. The only resultative
symmetry transformations are given by the matrix ${\cal C}_R^2$, i.e. the string
theory extension of the stationary Einstein-Maxwell theory is three-parametric.

Let us now calculate this extension. Let us put $\lambda^2_{\mu}=\lambda_{\mu}$
for simplicity and apply Eqs. (\ref{e3}), (\ref{e9}), (\ref{e13}) and (\ref{e15}).
The result reads:
\be\label{e23}
{\cal Z}_1\!=\!(z_1^{'}\cosh{\lambda}\!+\!z_1^{''}\tilde\lambda_3)\!+\!
(z_2^{'}\tilde\lambda_1\!-\!z_2^{''}\tilde\lambda_2)\sigma_1\!+\!
(z_2^{''}\tilde\lambda_1\!+\!z_2^{'}\tilde\lambda_2)\sigma_3\!-\!
(z_1^{''}\cosh{\lambda}\!-\!z_1^{'}\tilde\lambda_3)\epsilon,
\ee
where $\tilde\lambda_{\mu}=\lambda^{-1}\sinh{(\lambda)}\,\lambda_{\mu}$,
whereas ${\cal Z}_2$ can be obtained from ${\cal Z}_1$ using the substitution
$z_1\leftrightarrow z_2$. In Appendix B one can find a material related to
calculation of the Ernst matrix potentials and a (partial) encoding of the
$\sigma$-model information the language of the physical field components.


\section{Generation using complex potentials}
\setcounter{equation}{0}

Let us consider the $2\times 2$ nonconstrained complex matrix ${\bf Z}$; it has
the same number of
degrees of freedom as the potential ${\cal Z}$. Let also ${\bf C}_L$ is the $U(1)$
group factor and both the $2\times 2$ matrices ${\bf C}_R^1$ and ${\bf C}_R^2$
parametrizes the group $SU(1,1)$. Then the transformation
\be\label{e24}
{\bf Z}\rightarrow {\bf C}_L{\bf C}_R^{2\,T}{\bf Z}{\bf C}_R^1
\ee
realizes the $U(1)\times SU(1,1)\times SU(1,1)$ symmetry in action on the
matrix ${\bf Z}$.
We would like to use the group isomorphisms $U(1)\sim SO(2)$ and
$SU(1,1)\sim SO(2,1)$ to parametrize the matrix ${\bf Z}$ by the components of
${\cal Z}$ in such a way when Eq. (\ref{e24}) will be a complex equivalent of Eq.
(\ref{e3}). In the case of realization of this program we will have a formalism
of the complex matrix potential which transforms linearly under the action of the
charging symmetry subgroup of transformations. This new formalism will be the
most compact and actually convenient for generation of asymptotically flat
solutions of the theory. Moreover, it will be more natural than the real one in
respect to string theory symmetry extension of the stationary Einstein-Maxwell
theory which is naturally parametrized by the complex potential ${\bf z}$
\, (see Eq. \ref{e17}).

Our plan is the following: we will calculate the functional representation for the
generators $`1'$ and $`2'$ in real ($\Gamma_1^a$ and $\Gamma_2^a$ in terms of the
${\cal Z}$-components) and complex (${\bf\Gamma}_1^a$ and ${\bf\Gamma}_2^a$ in terms
of the ${\bf Z}$-components) variables and will identify the corresponding
generators using the proposing functional dependence between the real and complex
variables. After that we will demand a proportionality $\Gamma_0\sim{\bf\Gamma}_0$
for utilization of the $U(1)$ symmetry. The last step will be related with the
convenient fixation of `gauge' in the obtained solution - the explicit
parametrization of ${\bf Z}$ in terms of ${\cal Z}$ components. Of course,
we will not additionally consider the equality $\Gamma_3^a={\bf\Gamma}_3^a$,
because it will authomatically take place in view of the previous steps and the
commutation relations.

In order to realize this plan let us parametrize the potentials $\bf Z$ and ${\cal Z}$
as
\be\label{e25}
{\bf Z}=\left (\ba{cc}
z_1&z_3\cr
z_4&z_2
\ea\right ),\quad
{\cal Z}=\left (\ba{cccc}
\zeta_1&\zeta_3&\zeta_5&\zeta_7\cr
\zeta_4&\zeta_2&\zeta_8&\zeta_6
\ea\right )
\ee
and calculate the generators in both the representations. From Eq. (\ref{e3})
it follows that the infinithesimal transformations of the potential ${\cal Z}$
read:
\be\label{e26}
\delta_{\mu}^a{\cal Z}&=&{\cal Z}\Gamma^a_{\mu},\quad \delta_0{\cal Z}=
\Gamma_0{\cal Z},
\ee
so by the help of Eq. (\ref{e25}) one obtains for the real form of the
functional generators the following expressions:
\be\label{e27}
\Gamma_1^1&=&\zeta_5\frac{\partial}{\partial\zeta_1}+\zeta_1\frac{\partial}{\partial\zeta_5}
+\zeta_6\frac{\partial}{\partial\zeta_2}+\zeta_2\frac{\partial}{\partial\zeta_6}
+\zeta_3\frac{\partial}{\partial\zeta_7}+\zeta_7\frac{\partial}{\partial\zeta_3}
+\zeta_4\frac{\partial}{\partial\zeta_8}+\zeta_8\frac{\partial}{\partial\zeta_4},
\nonumber\\
\Gamma_2^1&=&-\zeta_7\frac{\partial}{\partial\zeta_1}-\zeta_1\frac{\partial}{\partial\zeta_7}
+\zeta_2\frac{\partial}{\partial\zeta_8}+\zeta_8\frac{\partial}{\partial\zeta_2}
-\zeta_4\frac{\partial}{\partial\zeta_6}-\zeta_6\frac{\partial}{\partial\zeta_4}
+\zeta_5\frac{\partial}{\partial\zeta_3}+\zeta_3\frac{\partial}{\partial\zeta_5};
\nonumber\\
\Gamma_1^2&=&\zeta_7\frac{\partial}{\partial\zeta_1}+\zeta_1\frac{\partial}{\partial\zeta_7}
+\zeta_8\frac{\partial}{\partial\zeta_2}+\zeta_2\frac{\partial}{\partial\zeta_8}
+\zeta_5\frac{\partial}{\partial\zeta_3}+\zeta_3\frac{\partial}{\partial\zeta_5}
+\zeta_6\frac{\partial}{\partial\zeta_4}+\zeta_4\frac{\partial}{\partial\zeta_6},
\nonumber\\
\Gamma_2^2&=&\zeta_5\frac{\partial}{\partial\zeta_1}+\zeta_1\frac{\partial}{\partial\zeta_5}
-\zeta_7\frac{\partial}{\partial\zeta_3}-\zeta_3\frac{\partial}{\partial\zeta_7}
+\zeta_8\frac{\partial}{\partial\zeta_4}+\zeta_4\frac{\partial}{\partial\zeta_8}
-\zeta_6\frac{\partial}{\partial\zeta_2}-\zeta_2\frac{\partial}{\partial\zeta_6};
\nonumber\\
\Gamma_0&=&\zeta_1\frac{\partial}{\partial\zeta_4}-\zeta_4\frac{\partial}{\partial\zeta_1}
+\zeta_3\frac{\partial}{\partial\zeta_2}-\zeta_2\frac{\partial}{\partial\zeta_3}
+\zeta_5\frac{\partial}{\partial\zeta_8}-\zeta_8\frac{\partial}{\partial\zeta_5}
+\zeta_7\frac{\partial}{\partial\zeta_6}-\zeta_6\frac{\partial}{\partial\zeta_7}.
\ee
Then, from Eq. (\ref{e24}) for the infinithesimal transformations of the potential
${\bf Z}$ one obtains:
\be\label{e28}
{\bf\delta}_{\mu}^1{\bf Z}={\bf Z}{\bf\Gamma}_{\mu}^1,\quad
{\bf\delta}_{\mu}^2{\bf Z}={\bf\Gamma}_{\mu}^{2\,T}{\bf Z},\quad
{\bf\delta}_0{\bf Z}={\bf\Gamma}_0{\bf Z},
\ee
and, in view of Eq. (\ref{e25}),  the corresponding complex generators read:
\be\label{e29}
{\bf\Gamma}_1^1&=&z_3\frac{\partial}{\partial z_1}+z_1\frac{\partial}{\partial z_3}
+z_4\frac{\partial}{\partial z_2}+z_2\frac{\partial}{\partial z_4},
\nonumber\\
{\bf\Gamma}_2^1&=&i\left ( z_3\frac{\partial}{\partial z_1}-z_1\frac{\partial}{\partial z_3}
+z_2\frac{\partial}{\partial z_4}-z_4\frac{\partial}{\partial z_2}\right );
\nonumber\\
{\bf\Gamma}_1^2&=&z_4\frac{\partial}{\partial z_1}+z_1\frac{\partial}{\partial z_4}
+z_3\frac{\partial}{\partial z_2}+z_2\frac{\partial}{\partial z_3},
\nonumber\\
{\bf\Gamma}_2^2&=&i\left ( z_4\frac{\partial}{\partial z_1}-z_1\frac{\partial}{\partial z_4}
+z_2\frac{\partial}{\partial z_3}-z_3\frac{\partial}{\partial z_2}\right );
\nonumber\\
{\bf\Gamma}_0&=&i\left ( z_1\frac{\partial}{\partial z_1}+z_2\frac{\partial}{\partial z_2}
+z_3\frac{\partial}{\partial z_3}+z_4\frac{\partial}{\partial z_4}\right ),
\ee
where only the holomorphic part is presented. For our purposes it will be convenient to
have a linear ${\bf Z}$-${\cal Z}$ dependence. In fact it is possible to search it
using an ansatz $z_{i'}={\cal D}_{i'i''}\zeta_{i''}$, where $i'=1,...4$ and
$i''=1,...8$. The process of identification of the corresponding generators is trivial, it
leads to establishing of different relations between the matrix ${\cal D}$ components.
The resulting parametrization of ${\bf Z}$ in terms of ${\cal Z}$ components can
be represented in the following form: let us introduce the quantities
\be\label{e30}
\hat z_1&=&\zeta_1+i\zeta_3-\sigma\left (\zeta_2-i\zeta_4\right ),
\nonumber\\
\hat z_2&=&\zeta_1-i\zeta_3+\sigma\left (\zeta_2+i\zeta_4\right ),
\nonumber\\
\hat z_3&=&\zeta_5+i\zeta_7-\sigma\left (\zeta_6-i\zeta_8\right ),
\nonumber\\
\hat z_4&=&\zeta_5-i\zeta_7+\sigma\left (\zeta_6+i\zeta_8\right ),
\ee
where the real constant $\sigma$ is the coefficient in the equality
$\Gamma_0=\sigma{\bf\Gamma}_0$. Then all the generators become identified
if $\sigma=\pm 1$ and
\be\label{e31}
{\bf Z}=\left (\ba{cc}
{\cal D}_{11}\hat z_1+{\cal D}_{15}\hat z_4&
{\cal D}_{11}\hat z_3+{\cal D}_{15}\hat z_2\cr
{\cal D}_{21}\hat z_4+{\cal D}_{25}\hat z_1&
{\cal D}_{21}\hat z_2+{\cal D}_{25}\hat z_3
\ea\right ).
\ee
Here the parameters ${\cal D}_{11},{\cal D}_{15},{\cal D}_{21},{\cal D}_{25}$ are
restricted only by the demanding that the functional dependence of $z_{i'},\bar z_{i''}$
on $\zeta_{i''}$ must be nondegenerated. In fact the freedom in choice of these
parameters can be considered as some gauge freedom. Actually, let us consider the
transformation
\be\label{e32}
{\bf Z}\rightarrow {\bf L}{\bf Z}{\bf R},
\ee
with the nondegenerated constant matrices ${\bf L}$ and ${\bf R}$.
It induces the following transformations of the group matrices of the following form:
\be\label{e33}
{\bf C}_R^1\rightarrow {\bf R}{\bf C}_R^1{\bf R}^{-1},\quad
{\bf C}_R^2\rightarrow {\bf L}^T{\bf C}_R^2{\bf L}^{-1,\,T}.
\ee
Such transformations do not change the underlying generator multiplication
relations and leave our scheme of generator identifications unchanged. It is easy
to prove that it is possible to take the matrices $\bf L$ and $\bf R$ in such a
way that finally $z_{i'}=\hat z_{i'}/2$. Also it is convenient to take $\sigma=-1$.
Then
\be\label{e34}
{\bf Z}=\frac{1}{2}\left (\ba{cc}
\zeta_1+\zeta_2+i(\zeta_3-\zeta_4)&\zeta_5+\zeta_6+i(\zeta_7-\zeta_8)
\cr
\zeta_5-\zeta_6-i(\zeta_7+\zeta_8)&\zeta_1-\zeta_2-i(\zeta_3+\zeta_4)
\ea\right ).
\ee

Now let us consider the generation starting from the stationary Einstein-Maxwell
fields. In this case (see Eqs. (\ref{e15}), (\ref{e25}))
\be\label{e35}
{\bf Z}_{EM}=\left (\ba{c}
z
\cr
0
\ea\right ),
\ee
where the $1\times 2$ matrix $z$ iz the Einstein-Maxwell potential from
Eq. (\ref{e17'}). Taking into account the reasons given in the previous
section and omit ${\bf C}_L$ and ${\bf C}_R^1$ transformations and also putting
${\bf C}_R^{2}={\bf C}$, one obtains that
\be\label{e36}
{\bf Z}={\bf C}^T{\bf Z}_{EM},
\ee
where
\be\label{e37}
{\bf C}=\cosh{\lambda}+\frac{\sinh{\lambda}}{\lambda}\hat{\bf\lambda},
\ee
and $\hat{\bf\lambda}=\lambda_{\mu}{\bf \Gamma}_{\mu}$. Equations
(\ref{e34})-(\ref{e37}) completely realize the scheme of generation of
the string theory solutions starting from the stationary Einstein-Maxwell
fields in framework of the complex matrix potential formalism.


\section*{Acknowledgments}
This work was supported by RFBR grant ${\rm N^{0}}
\,\, 00\,02\,17135$.


\section{Conclusion}
\setcounter{equation}{0}

In this article it is developed a formalism for generation of the
stationary solutions of four-dimensional low-energy heterotic string
theory with two Maxwell fields starting from the stationary Einstein-Maxwell
theory. This formalism is the most general in respect to the stationary
charging symmetry subgroup; it realizes the $SU(1,1)$-covariant string theory
extension
of the charging symmetry complete Einstein-Maxwell stationary solution space.
Both the $2\times 4$ real and $2\times 2$ complex matrix potential based
approaches
are developed; a relation between them is given by Eq. (\ref{e34}). In the
forthcoming publications I hope to generalize some results to the general case
of arbitrary value of $d$ and $n$ of the heterotic string gravity model
(\ref{e1}), and also to
give some explicit examples of generation of the concrete solutions. Here it
is useful only to stress that the new generation technique immediately gives
the
action of charging symmetries on the charges of asymptotically flat solutions.
Actually, it is easy to prove that the potential
\be\label{e38}
z=\frac{q}{R}
\ee
where $q$ is the complex constant $1\times 2$ matrix, together with
the 3-dimensional line element
\be\label{e39}
ds_3^2=dR^2+\left(R^2-I\right)d\Omega^2,
\ee
where $I=qq^+$, is the solution of motion equations for the system (\ref{e17}).
This asymptotically flat
solution is in fact the Reissner-Nordstrom one (see \cite{EMT}), and the matrix
$q$ components can be easily related to the mass, parameter NUT and also
to electric and
magnetic charges of this point-like source. It is easy to see that the
transformations ${\bf C}_L$ and ${\bf C}_R^1$ provide the pure
$q$-reparametrizations and can actually be omitted without loss of any
generality. The remaining `second' $SU(1,1)$ transformation (we use the
language of complex representation for definiteness) acts as
\be\label{e40}
{\bf Q}\rightarrow{\bf C}^T{\bf Q}
\ee
on the charge matrix of the solution, defined accordingly the relation
\be\label{e41}
{\bf Z}=\frac{{\bf Q}}{R}.
\ee
In this case Eq. (\ref{e41}) gives exact value of the ${\bf Z}$-potential. In the
general case of asymptotically flat solution Eq. (\ref{e41}) defines the monopole
term which is nonzero for the charged solutions. In this general situation Eq.
(\ref{e40}) preserves its previous exact sense of the heterotic string theory
generalization of the effective Einstein-Maxwell theory system of charges.

At the end of our discussion let us note that using the developed generation
technique one obtains only some subspace of the stationary solution space of
the theory under consideration. Actually, in general situation the matrix potential
${\bf Z}$ is a complex matrix field free of any nondynamical restrictions.
However, for ${\bf Z}$ generated from ${\bf Z}_{EM}$ one obtains that
${\rm det}\,{\bf Z}=0$ identically, as it immediately follows from Eq. (\ref{e36}).
The same conclusion can be done about the generality of charge characteristics
of the generated solution. This opportunity seems interesting in the context
of concrete work when often the real number of independent parameters of the
constructed solution remains hidden in complicated notations. Our formalism
guarantees clear form of all results in view of its explicit charging symmetry
covariance.

\renewcommand{\theequation}{A.\arabic{equation}}
\section*{Appendix A}
\setcounter{equation}{0}

The components of the original multidimensional fields can be embedded into
three groups. The first group consists of the 3-dimensional scalars; these
are the matrices $G$, $B$ and $A$ with the components $G_{mk}$, $B_{mk}$
and $A_{mI}=A_m^I$.  Their dimensions are $d\times d$, $d\times d$ and
$d\times n$ respectively. Also there is a scalar field
\be\label{A0}
\phi=\Phi-{\rm ln}\sqrt{-{\rm det}G}.
\ee
The second group contains 3-vectors columns
$\vec A_1$, $\vec A_2$ and $\vec A_3$ of the dimension $d\times 1$, $d\times 1$
and $n\times 1$ respectively. Their components read:
\be\label{A1}
\left (\vec A_1\right )_{m\mu}&=&\left ( G^{-1}\right )_{mk}G_{k,d+\mu},
\nonumber\\
\left ( \vec A_2\right )_{m\mu}&=&B_{m,d+\mu}-B_{mn}\left ( \vec A_1\right )_
{n\mu}
+\frac{1}{2}A_{mI}\left (\vec A_3\right )_{I\mu},
\nonumber\\
\left ( \vec A_3\right )_{I\mu}&=&-A^I_{d+\mu}+A_{mI}\left ( \vec A_1\right )_
{m\mu}.
\ee
The third group is the group of 3-dimensional tensor fields; there are two
ones:
\be\label{A2}
h_{\mu\nu}&=&e^{-2\phi}\left [ G_{\mu\nu}-G_{mk}\left ( \vec A_1\right )_{m\mu}
\left ( \vec A_1\right )_{k\nu}\right ],
\nonumber\\
b_{\mu\nu}&=&B_{\mu\nu}-B_{mk}\left ( \vec A_1\right )_{m\mu}
\left ( \vec A_1\right )_{k\nu}-\frac{1}{2}\left [
\left ( \vec A_1\right )_{m\mu}\left ( \vec A_2\right )_{m\nu}-
\left ( \vec A_1\right )_{m\nu}\left ( \vec A_2\right )_{m\mu}
\right ].
\ee
Following \cite{Sen3} we put $b_{\mu\nu}=0$. Then, using the motion equations it
is possible to `dualize' the vector fields; let $u$, $v$ and $s$ are the
$d\times 1$, $d\times 1$ and $d\times n$ (pseudo)scalar 3-fields related to the
corresponding vector ones through the relations
\be\label{A3}
\nabla\times\vec A_1&=&e^{2\phi}G^{-1}\left [ \nabla u+
\left ( B+\frac{1}{2}AA^T
\right )\nabla v+A\nabla s\right ],
\nonumber\\
\nabla\times\vec A_2&=&e^{2\phi}G\nabla v-\left ( B+\frac{1}{2}AA^T\right )
\nabla\times\vec A_1+A\nabla\times\vec A_3,
\nonumber\\
\nabla\times\vec A_3&=&e^{2\phi}\left ( \nabla s+A^T\nabla v\right )+
A^T\nabla\times\vec A_1.
\ee
Finally, the set of the effective 3-dimensional scalars and pseudoscalars
contains the quantities
$G$, $B$, $A$, $\phi$, $u$, $v$ and $s$. They parametrize the
heterotic string theory induced nonlinear $\sigma$-model coupled to gravity.

The Ernst matrix potential formulation is based on the use of two matrices
\be\label{A4}
{\cal X}=\left (
\ba{cc}
-e^{-2\phi}+v^TXv+v^TAs+1/2\,s^Ts&v^TX-u^T\cr
Xv+u+As&X
\ea\right ),
\quad
{\cal A}=\left (\ba{c}
s^T+v^TA\cr
A
\ea\right ),
\ee
where $X=G+B+1/2\,AA^T$ (\cite{HK1},\cite{HK2}); in terms of them the
effective 3-dimensional action reads:
\be\label{A5}
S_3\!=\!\int dx \sqrt h \left\{\!-\!R_3\!+\!{\rm Tr}\left [
\frac{1}{4}\left (\nabla{\cal X}\!-\!\nabla{\cal A}{\cal A}^T\right )
{\cal G}^{-1}
\left (\nabla{\cal X}^T\!-\!{\cal A}\nabla{\cal A}^T\right ){\cal G}^{-1}+
\frac{1}{2}\nabla{\cal A}^T{\cal G}^{-1}{\nabla A}\right ]\right\},
\nonumber\\
\ee
where
\be\label{A5'}
{\cal G}=1/2 ({\cal X}+{\cal X}^T-{\cal A}{\cal A}^T).
\ee
The null-curvature matrix formulation is related to the matrix ${\cal M}$,
which can be defined using the Ernst matrix potentials \cite{OK}:
\be\label{A6}
{\cal M}=\left (\ba{ccc}
{\cal G}^{-1}&{\cal G}^{-1}{\cal X}-1&{\cal G}^{-1}{\cal A}\cr
{\cal X}^T{\cal G}^{-1}-1&{\cal X}^T{\cal G}^{-1}{\cal X}&
{\cal X}^T{\cal G}^{-1}{\cal A}\cr
{\cal A}^T{\cal G}^{-1}&{\cal A}^T{\cal G}^{-1}{\cal X}&
{\cal A}^T{\cal G}^{-1}{\cal A}+1
\ea\right ).
\ee
This matrix parametrizes the coset $O(d+1,d+1+n)/O(d+1)\times O(d+1+n)$ in view
of the number of its independent components and the identical satisfaction of
the relations
\be\label{A7}
{\cal M}^T={\cal M}, \quad {\cal M}{\cal L}{\cal M}={\cal L},
\ee
where
\be\label{A8}
{\cal L}=\left (\ba{ccc}
0&1&0\cr
1&0&0\cr
0&0&-1
\ea\right ).
\ee
In terms of the curvature matrix the 3-action takes the following form:
\be\label{A9}
S_3=\int dx \sqrt h \left\{-R_3+\frac{1}{8}{\rm Tr}\left [
\left (\nabla{\cal M}{\cal M}^{-1}\right )^2
\right ]\right\}.
\ee
The ${\cal Z}$-potential formulation can also be defined through the
Ernst matrix potentials. To do it, let us define the matrices
\be\label{A10}
{\cal Z}_1=2\left ( {\cal X}+\Sigma\right )^{-1}-\Sigma,\quad
{\cal Z}_2=\sqrt 2\left ( {\cal X}+\Sigma\right )^{-1}{\cal A}.
\ee
Then
\be\label{A11}
{\cal Z}=\left ( {\cal Z}_1\,\,{\cal Z}_2\right ).
\ee
The ${\cal Z}$-expressed effective $3$-dimensional action is given
by Eq. (\ref{e2}).


\section*{Appendix B}
\renewcommand{\theequation}{B.\arabic{equation}}
\setcounter{equation}{0}

It is easy to verify that the matrices
\be\label{B1}
g_0=1,\quad g_1=\sigma_1, \quad g_2=\sigma_3, \quad g_3=-\epsilon
\ee
satisfy the same multiplicative relations as the in Eq. (\ref{e11}).
In fact the set $\{g_{\aleph}\} \,\, (\aleph=0,...,3)$ give the
generators of the $sl(2,R)$
algebra which is isomorphic to $so(1,2)$. The multiplication table can
be used for the explicit computation of the Ernst matrix potentials
and for the following obtaining of components of the
physical fields. In this computation it will be convenient to use the
decompositions of all the $2\times 2$ matrices in respect to the
basis (\ref{B1}) (for example, the decomposition
${\cal Z}^a={\cal Z}^a_{,\aleph}\,g_{\aleph}$ is given in
Eq. (\ref{e23})).

In computation of the Ernst matrix potentials one must take into
account their $2\times 2$ matrix dimensionality; from this fact and
Eq. (\ref{A10}) it follows that
\be\label{B2}
{\cal X}=-1+\frac{2}{\Delta}\left ( -{\cal Z}_1+{\rm det}{\cal Z}_1
\right ), \quad
{\cal A}=-\frac{\sqrt 2}{\Delta}
\left ( 1-{\cal Z}_1^{\star}\right ){\cal Z}_2,
\ee
where $\Delta=1-{\rm Tr}{\cal Z}_1+{\rm det}{\cal Z}_1$
and ${\cal Z}_1^{\star}=-\epsilon{\cal Z}_1^T\epsilon=
{\cal Z}_{1,0}g_0-{\cal Z}_{1,\mu}g_{1,\mu}$. Here
${\cal Z}_1^{\star}{\cal Z}_2=({\cal Z}_1^{\star}{\cal Z}_2)_{,\aleph}
g_{\aleph}$, where
\be\label{B3}
({\cal Z}_1^{\star}{\cal Z}_2)_{,0}&=&
{\cal Z}_{1,0}{\cal Z}_{2,0}-{\cal Z}_{1,1}{\cal Z}_{2,1}-
{\cal Z}_{1,2}{\cal Z}_{2,2}+{\cal Z}_{1,3}{\cal Z}_{2,3},
\nonumber\\
({\cal Z}_1^{\star}{\cal Z}_2)_{,1}&=&
{\cal Z}_{1,0}{\cal Z}_{2,1}-{\cal Z}_{2,0}{\cal Z}_{1,1}+
{\cal Z}_{1,2}{\cal Z}_{2,3}-{\cal Z}_{1,3}{\cal Z}_{2,2},
\nonumber\\
({\cal Z}_1^{\star}{\cal Z}_2)_{,2}&=&
{\cal Z}_{1,0}{\cal Z}_{2,2}-{\cal Z}_{2,0}{\cal Z}_{1,2}+
{\cal Z}_{1,3}{\cal Z}_{2,1}-{\cal Z}_{1,1}{\cal Z}_{2,3},
\nonumber\\
({\cal Z}_1^{\star}{\cal Z}_2)_{,3}&=&
{\cal Z}_{1,0}{\cal Z}_{2,3}-{\cal Z}_{2,0}{\cal Z}_{1,3}-
{\cal Z}_{1,1}{\cal Z}_{2,2}+{\cal Z}_{1,2}{\cal Z}_{2,1}.
\ee
Also it is easy to see that ${\rm Tr}{\cal Z}_1=2{\cal Z}_{1,0}$ and
${\rm det}{\cal Z}_1={\cal Z}_{1,0}^2-{\cal Z}_{1,1}^2-{\cal Z}_{1,2}^2
+{\cal Z}_{1,3}^2$.
Then, using definitions of the potentials ${\cal X}$ and ${\cal A}$
(see Eq. (\ref{A4})), one can
calculate all the scalar and pseudoscalar variables. For example, for
the electric potentials one obtains:
\be\label{B4}
A^1_t&=&-\frac{\sqrt 2}{\Delta}\left [
{\cal Z}_{2,1}+{\cal Z}_{2,3}-
\left ({\cal Z}_1^{\star}{\cal Z}_2\right )_{,1}-
\left ({\cal Z}_1^{\star}{\cal Z}_2\right )_{,3}
\right ],
\nonumber\\
A^2_t&=&-\frac{\sqrt 2}{\Delta}\left [
{\cal Z}_{2,0}-{\cal Z}_{2,2}-
\left ({\cal Z}_1^{\star}{\cal Z}_2\right )_{,0}+
\left ({\cal Z}_1^{\star}{\cal Z}_2\right )_{,2}
\right ].
\ee
Then, calculating ${\cal G}$ accordingly Eq. (\ref{A5'})
by exploring the multiplication table (\ref{e11}), one
obtains the remaining nondualizing components.
Here we will not give them here in view of their complicated
form as well as the expressions for the redualized 3-vectors
and the related physical field components. They will be
presented in the following publications related to the concrete
generations of new solutions of low-energy heterotic string
theory.


\end{document}